\documentclass[conference]{IEEEtran}
\IEEEoverridecommandlockouts
\usepackage{cite}
\usepackage{amsmath,amssymb,amsfonts}
\usepackage{algorithmic}
\usepackage{graphicx}
\usepackage{textcomp}
\usepackage{xcolor}
\usepackage{url}
\usepackage{hyperref}
\def\BibTeX{{\rm B\kern-.05em{\sc i\kern-.025em b}\kern-.08em
    T\kern-.1667em\lower.7ex\hbox{E}\kern-.125emX}}

\begin{document}

\title{SYNAPSE: A Multi-LLM Orchestrated AI Tutor for Secure Software Development Education with Neurodivergent-First Design\thanks{© 2026 IEEE. This is the accepted version of a paper accepted for publication at the 2026 IEEE International Conference on Software Maintenance and Evolution (ICSME), licensed under CC BY 4.0.}}

\author{
\IEEEauthorblockN{Giusy Ferrara \quad\quad Ashkan Sami}
\IEEEauthorblockA{School of Computing, Engineering \& The Built Environment\\
Edinburgh Napier University, Edinburgh, United Kingdom\\
giusyferrara@live.com \quad\quad A.Sami@napier.ac.uk}
}

\maketitle

\begin{abstract}
Developers who maintain real systems must continually recognise and
remediate vulnerabilities in existing code, yet this skill is rarely
trained directly: secure software development is commonly taught only
after programming fluency is acquired, and accessibility support is
treated as a secondary concern, disadvantaging learners with ADHD and
related executive-function differences.
This paper presents SYNAPSE, a publicly deployed adaptive tutoring
platform for Java, Python for cybersecurity, and secure software
development. SYNAPSE coordinates Claude, GPT-4o, and Gemini through
the Model Context Protocol, routing interactions by pedagogical
intent under a three-stage Socratic hint policy. It exposes eighteen
always-visible accessibility features and anchors practice in
ShopSecure, a deliberately vulnerable web application mapped to six
OWASP Top~10 (2021) categories, on which learners practise the
detect--understand--remediate loop characteristic of software
maintenance.
A feasibility pilot with nineteen participants across neurodivergent
and neurotypical cohorts returned a System Usability Scale score of
76.4 and engagement of 4.2/5, with comparable cognitive-load levels
across cohorts. SYNAPSE is available at https://synapse-course.com;
a screencast is available at https://youtu.be/9R17KC47qQI.
\end{abstract}
\begin{IEEEkeywords}
software maintenance and evolution, vulnerability remediation,
AI-assisted programming education, Model Context Protocol,
secure software development, neurodivergent learners
\end{IEEEkeywords}
\section{Introduction}

Modern software underpins healthcare, finance, transportation, and critical infrastructure, and most of its lifetime is spent being maintained and evolved rather than written from scratch. A large share of that maintenance effort goes into recognising, understanding, and remediating vulnerabilities in existing code; many vulnerabilities can be viewed as a form of security debt that must be identified and addressed during maintenance. The skill this work demands is therefore not greenfield secure coding but the detect--understand--remediate loop applied to code that already exists. Yet two structural patterns in computing education limit how well developers are prepared for it.

The first is the separation between programming and security. Most curricula and self-paced platforms (Codecademy, TryHackMe, HackTheBox) teach programming fluency first and introduce secure software development only afterwards, by which point insecure habits have already formed \cite{allodi2020economics, votipka2018hackers}. Crucially, vulnerability assessment and repair are maintenance activities performed on existing systems: Allodi et al.~\cite{allodi2020economics} show that targeted training improves vulnerability assessment but that learner accuracy still lags confidence, and Votipka et al. \cite{votipka2018hackers} show that success in vulnerability discovery depends on exposure to realistic system-level scenarios that introductory curricula defer. The rise of AI coding assistants sharpens this concern: Belozerov et al.~\cite{belozerov2026securityllm} evaluate contemporary models across 2,315 code snippets and find that even strong models confidently miss or mis-repair real vulnerabilities, a detection-to-repair failure that lands squarely in the maintenance phase.

The second is a less visible separation: mainstream platforms assume a neurotypical learner profile. Le Cunff et al. \cite{lecunff2024neurodivergent} report that the large majority of online-learning studies do not treat neurodiversity as a measurable variable, leaving learners with ADHD and related executive-function differences largely invisible in the evidence base used to design these environments.

At the same time, recent results suggest that carefully designed AI tutors can outperform conventional active-learning approaches in both outcomes and engagement \cite{kestin2025aitutor}. The opportunity is real, but so is the documented risk: unstructured AI assistance erodes critical thinking~\cite{zhai2024effects} and does not transfer to independent tasks \cite{bastani2025genai}. In security specifically, even advanced models miss real vulnerabilities or generate confidently incorrect remediation \cite{belozerov2026securityllm}.

This paper presents SYNAPSE, a publicly deployed adaptive tutoring platform that addresses these gaps as a single design problem and frames secure-coding competence as a maintenance competence. SYNAPSE teaches Java programming and secure software development as inseparable concerns from the first module, training the recognition-and-remediation skill that maintenance work demands; it exposes an integrated accessibility layer designed for ADHD and executive-function needs; and it coordinates three large language models (Claude, GPT-4o, and Gemini) through the Model Context Protocol (MCP) under a Socratic, scaffolded hint policy. The platform is available at \url{https://synapse-course.com}.

The contributions of this paper are:
\begin{itemize}
\item A working, publicly accessible tool integrating multi-LLM orchestration via MCP with a neurodivergent-first interface.
\item A three-stage Socratic hint policy enforced across providers to mitigate over-reliance on direct answers.
\item ShopSecure, a purpose-built, deliberately vulnerable Flask application bundled with the platform and mapped to six OWASP Top 10 (2021) categories, a reusable corpus of vulnerability patterns in a working application on which the detect--understand--remediate loop can be practised in-browser.
\item Pilot evaluation evidence from nineteen participants (mixed neurodivergent and neurotypical), reporting SUS 76.4, engagement 4.2/5, and correct mitigation selection across CWE-22, CWE-352, and CWE-502 among security-pathway completers, presented as feasibility evidence.
\end{itemize}

\section{Background and Related Work}
\label{sec:background}
Research-designed AI tutors can outperform conventional active
learning~\cite{kestin2025aitutor}, yet unstructured assistance may
erode critical thinking~\cite{zhai2024effects} and fail to transfer
to independent tasks~\cite{bastani2025genai}. In programming
education, Socratic and scaffolded guidance preserves learner
agency~\cite{kazemitabaar2023studying, elnaffar2026teaching},
indicating that hints should be released progressively rather than
as direct answers.

For neurodivergent learners, the FEDIS+R
framework~\cite{lecunff2025fedis} structures cognitive-load
management around format, environment, delivery, instruction, and
support, but remains a tentative model untested in deployed AI
tutoring systems, while neurodiversity is underrepresented in the
online-learning evidence base~\cite{lecunff2024neurodivergent}.

In secure development, education (not experience alone) predicts
vulnerability-assessment accuracy~\cite{allodi2020economics}, and
realistic system-level practice is essential for discovery
skills~\cite{votipka2018hackers}. Single LLMs are unreliable here:
Belozerov et al.~\cite{belozerov2026securityllm} found that no
individual model detected all vulnerabilities in real-world code,
that different models missed different weaknesses, and that all
produced confidently incorrect diagnoses. Orchestrating multiple
models by pedagogical role therefore mitigates single-model blind
spots.

Prior work addresses AI tutoring, accessibility, and secure-coding
practice in isolation; to our knowledge, no existing tool combines scaffolded
multi-LLM tutoring with accessibility-first design for vulnerability
recognition and remediation. This gap motivates SYNAPSE.

\section{SYNAPSE Architecture}
\label{sec:architecture}

\begin{figure*}[t]
  \centering
  \includegraphics[width=\textwidth]{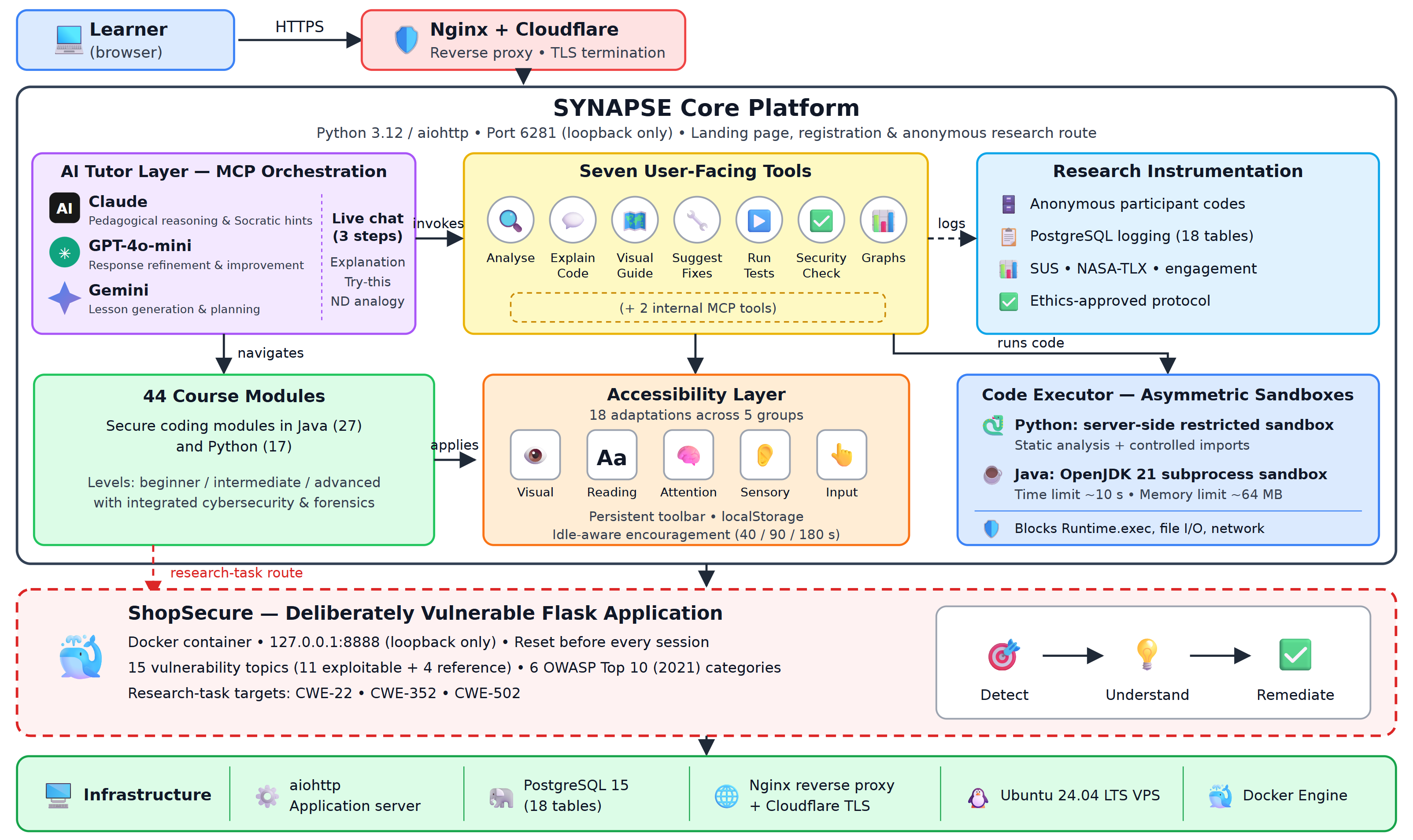}
  \caption{SYNAPSE high-level architecture. The red dashed arrow
  marks the research-task route to ShopSecure, which runs in an
  isolated Docker container.  }
  \label{fig:__architecture}
\end{figure*}

SYNAPSE is a publicly deployed web platform built around five
interacting subsystems: an adaptive AI tutor coordinated by a Model
Context Protocol (MCP) server, a course content engine serving
forty-four structured modules, a persistent accessibility layer, a
dual-language code-execution environment, and \textit{ShopSecure}, a
deliberately vulnerable web application used as the hands-on security
target. A research instrumentation layer captures behavioural signals
under anonymous participant codes in a PostgreSQL database with
eighteen relational tables. The overall architecture is illustrated
in Fig.~\ref{fig:__architecture}.
\subsection{Multi-LLM Coordination via MCP}

The pedagogical core of SYNAPSE is an MCP server~\cite{anthropic2024mcp}
that exposes seven AI-mediated operations as callable tools:
\textit{Analyse and Learn}, \textit{Explain Code}, \textit{Visual Guide},
\textit{Suggest Fixes}, \textit{Run Tests}, \textit{Security Check},
and a \textit{Graphs and Visualisation} panel. Each tool maps a learner
intent (understanding, debugging, visualising) to a concrete dispatch
strategy. 

Learner-facing tutor responses are constructed through a three-step
composite pipeline. In \emph{Step~1}, Claude generates the core
explanation under a canonical \texttt{SYNAPSE\_SYSTEM\_PROMPT} that
enforces the Socratic interaction policy
(Section~\ref{subsec:socratic}). In \emph{Step~2}, GPT-4o-mini may
append a short practice exercise when the learner's query signals
educational intent. In \emph{Step~3}, Gemini~2.5~Flash may append a
learner-friendly explanatory analogy. Steps~2 and~3 are non-blocking:
if either provider fails, the response composed so far is returned,
ensuring graceful degradation. A separate branch routes requests
originating from the ShopSecure research task directly to Claude under
a task-calibrated prompt, bypassing the composite pipeline to maintain
alignment with vulnerability-discovery objectives.
The \textit{Explain Code} tool additionally uses Gemini as a planner
that identifies the language and recommends suitable pedagogical tools.

The use of three models reflects a division of pedagogical labour
rather than redundancy: each pipeline step assigns a distinct role
(Socratic explanation, practice-exercise generation, and analogy
construction) to a model selected according to the required interaction style, cost, and latency constraints. Proprietary hosted APIs were chosen over a
self-hosted open-weight model for pragmatic reasons: at this
project's scale, hosted inference provides the instruction-following
reliability required to enforce the hint policy without the
dedicated GPU infrastructure that a comparable open-weight
deployment would demand. The MCP architecture, however, keeps providers interchangeable:
an open-weight backend is a drop-in substitution rather than a
redesign. This also mitigates the behavioural drift across model
releases documented in~\cite{belozerov2026securityllm}. Whether multi-model
orchestration outperforms a well-designed single-model tutor remains
an empirical question this pilot was not designed to answer; we
identify a single-model ablation as future work.

\subsection{Three-Stage Socratic Hint Policy}
\label{subsec:socratic}

To mitigate the over-reliance risks documented
in~\cite{zhai2024effects,bastani2025genai}, all tutor responses are
constrained by a uniform Socratic hint policy enforced at the MCP
coordinator rather than within any single model. The policy organises
guidance into three progressively more specific stages:
(i)~\emph{conceptual reorientation}, which redirects the learner to
the underlying idea without naming the solution; (ii)~\emph{procedural
hint}, suggesting an approach or structural direction; and
(iii)~\emph{partial worked example} transferred from a different
context, requiring the learner to adapt the pattern. The policy
explicitly prohibits the generation of complete exploit code in
ShopSecure interactions. The same staged structure is reflected, at
the course-content layer, in three pre-written hints of increasing
specificity stored alongside each exercise, a design supported by
Kazemitabaar et al.~\cite{kazemitabaar2023studying}, who observed
learner dissatisfaction with AI systems that returned direct answers
instead of graduated guidance.
\begin{table}[!t]
    \centering
    \caption{The eighteen accessibility features of the SYNAPSE
    accessibility layer, grouped by function.}
    \label{tab:accessibility}
    \renewcommand{\arraystretch}{1.15}
    \begin{tabular}{p{0.28\columnwidth}p{0.60\columnwidth}}
        \hline
        \textbf{Group} & \textbf{Features} \\
        \hline
        Visual adjustments & Dark Mode, High Contrast, Dyslexia
        Font, Text Size, Line Spacing, Color Overlay \\
        Reading support & Reading Mask, Reading Guide, Read Aloud,
        Dictionary \\
        Attention \& focus & Focus Mode, Focus Assistant
        (idle-aware encouragement), Progress Bar, Cursor Highlight \\
        Sensory regulation & Calm Mode, Calming Audio, Reduce
        Motion \\
        Input alternatives & Voice Input \\
        \hline
    \end{tabular}
\end{table}
\subsection{Accessibility Layer}   
\label{subsec:accessibility}
The accessibility layer is available throughout the platform, ensuring
that support features remain accessible from any learning context. 
The eighteen features (Table~\ref{tab:accessibility}) are organised
into five groups and map onto the user-facing dimensions of the
FEDIS+R framework~\cite{lecunff2025fedis}. An idle-aware encouragement
system provides context-sensitive support during periods of inactivity,
while behavioural signals such as attention, cognitive load, and
emotional state are incorporated into tutor context construction. This
allows the AI tutor to adapt response length and complexity to learner
needs. Accessibility is therefore treated as a first-class pedagogical
component rather than a cosmetic interface layer.

\section{Implementation Highlights}
\label{sec:implementation}

\subsection{Technology Stack}

SYNAPSE is implemented in Python~3.12 on the \texttt{aiohttp}
asynchronous HTTP framework, chosen to support concurrent interactions
across multiple LLM providers. The MCP coordinator hosts seven
pedagogical tools behind a unified tool-listing and tool-calling
interface, while provider-specific clients for Anthropic, OpenAI, and
Google are encapsulated as thin wrappers. Persistent storage uses
PostgreSQL to manage research instrumentation, course content, and
per-learner progress. Behavioural instrumentation is recorded as raw
event logs without deriving or exposing interpreted learner metrics.

\subsection{ShopSecure: A Vulnerable Application for Hands-on Practice}
\textit{ShopSecure} is a Flask web application developed specifically
for SYNAPSE rather than a repackaging of existing vulnerable targets
such as OWASP Juice Shop~\cite{owasp_juice_shop} or
DVWA~\cite{dvwa}. Those platforms are standalone challenge
environments in which learners must first discover that a weakness
exists, without integrated tutoring. ShopSecure inverts this design:
vulnerabilities are overt, every page is wired into the Socratic AI
tutor with contextual hints and exploitation feedback, and the
pedagogical effort is spent on understanding and remediating each
weakness rather than on discovering it. It exposes fifteen
vulnerability topics (eleven actively exploitable endpoints plus four
reference code-pattern topics) covering six OWASP Top~10~(2021)
categories~\cite{owasp2021top10}.
\begin{table}[!t]
    \centering
    \caption{ShopSecure vulnerability inventory by OWASP Top~10~(2021) category. Research-task targets are marked $\star$.}
    \label{tab:vulnerabilities}
    \renewcommand{\arraystretch}{1.15}
    \begin{tabular}{p{0.42\columnwidth}p{0.40\columnwidth}c}
        \hline
        \textbf{OWASP Top~10 (2021)} & \textbf{CWEs Covered} & \textbf{\#} \\
        \hline
        A01 Broken Access Control & CWE-22$^{\star}$, CWE-352$^{\star}$, CWE-639 & 3 \\
        A03 Injection & CWE-78, CWE-79 (Refl.\ + Stored), CWE-89 ($\times 2$) & 5 \\
        A04 Insecure Design & CWE-209 & 1 \\
        A05 Security Misconfiguration & CWE-614 ($\times 2$) & 2 \\
        A07 Identification \& Auth.\ Failures & CWE-384, CWE-754, CWE-798 & 3 \\
        A08 SW \& Data Integrity Failures & CWE-502$^{\star}$ & 1 \\
        \hline
        \textbf{Total} & \textbf{6 categories} & \textbf{15} \\
        \hline
    \end{tabular}
\end{table}

The research task targets three of these, namely CWE-22 Path
Traversal, CWE-352 Cross-Site Request Forgery, and CWE-502
Insecure Deserialisation~\cite{mitre_cwe}. Each was chosen
because it exercises a distinct attacker capability: file-system
reach via parameter manipulation, authenticated cross-origin
coercion, and language-level remote code execution through
unsafe deserialisation. The most
pedagogically substantial scenario is a stored XSS-to-CSRF attack
chain, where the learner injects a hidden \texttt{<iframe>} into their
own profile so that an authenticated victim's browser silently issues
a transfer when viewing it, reframing CSRF as a realistic chained
exploit rather than a textbook single-vulnerability drill.

\subsection{Asymmetric Code Execution Sandbox}

Learner-submitted code is executed under an asymmetric
defence-in-depth strategy tailored to each language. Python code is
executed within a restricted sandbox using static analysis and
controlled imports, while Java code is compiled and executed as an
isolated OpenJDK~21 subprocess under resource limits. \textit{ShopSecure}
runs in a containerised environment separated from the tutoring
infrastructure, with each session restored to a consistent baseline.
The Python sandbox was hardened following an authorised pre-deployment
penetration test that identified an encoding-based privilege-escalation
path, reflecting the build--evaluate cycle of Design Science
Research~\cite{hevner2004designscience}.
\subsection{Usage Scenario: Remediating a Path-Traversal Defect}

To illustrate how SYNAPSE supports the detect--understand--remediate loop, we trace a learner through the CWE-22 exercise in ShopSecure, the platform's intended use: confronting a vulnerability in an existing, working application rather than writing secure code from scratch.

A beginner navigates to the \texttt{/invoice} endpoint, which the application surfaces directly. The home-page banner states plainly that this is a deliberately vulnerable application, so the task is never to discover that a weakness exists but to understand how it works and how to remediate it. The endpoint accepts a filename parameter, but the learner does not initially see the flaw.

Invoking the AI tutor, the request is routed through the MCP coordinator and, because it originates from the ShopSecure research task, dispatched directly to Claude under a task-calibrated prompt that bypasses the composite pipeline. The tutor responds under the three-stage Socratic hint policy: first a conceptual reorientation towards what the application does with the supplied filename, then, when the learner stalls, a procedural hint on how user input might reach the file system. The policy prohibits complete exploit code, guiding reasoning rather than handing over an attack string. Constructing a request that steps outside the intended directory, the learner triggers the visible ``PATH TRAVERSAL!'' banner, confirming understanding.

The learner then identifies the defect as CWE-22 and selects
an appropriate mitigation, completing the loop, as done by the
pilot participants who correctly selected mitigations for
CWE-22, CWE-352, and CWE-502.

\section{Pilot Evaluation}
\label{sec:evaluation}

\subsection{Study Design}

A feasibility pilot was conducted between February and April~2026
under Edinburgh Napier University ethics approval (Forms~C and~D).
Recruitment used informal convenience channels: peer student
networks, social media, and public online communities. Participation
was voluntary and unincentivized. Each participant accessed the
platform through an anonymous-coded research route, completed a
pre-task questionnaire, selected either the Java programming pathway
or the ShopSecure security challenge, and submitted a post-task
questionnaire combining the ten-item System Usability Scale
(SUS)~\cite{brooke1996sus}, an adapted three-scale NASA-TLX (Mental
Demand, Effort, Frustration)~\cite{hart1988nasatlx}, pre/post
confidence ratings, and three open-ended reflection items. By the
analysis cutoff of 15~April~2026, twenty-six volunteers had
registered and nineteen had completed a full session; the nineteen
completers form the analytic sample (seven self-identified
neurodivergent, twelve neurotypical; thirteen beginners and six
intermediate or advanced).

\subsection{Quantitative Results}

Table~\ref{tab:pilot-results} summarises the headline measures across
neurodivergent (ND) and neurotypical (NT) cohorts.

\begin{table}[!t]
    \centering
    \caption{Pilot study quantitative results across ND ($n=7$) and NT ($n=12$) cohorts. SUS, Engagement, and AI Helpfulness: higher is better. NASA-TLX scales: lower is better.}
    \label{tab:pilot-results}
    \renewcommand{\arraystretch}{1.15}
    \begin{tabular}{lccc}
        \hline
        \textbf{Measure} & \textbf{ND} & \textbf{NT} & \textbf{All} \\
        \hline
        SUS (0--100) & 76.8 & 76.1 & \textbf{76.4} \\
        Engagement (1--5) & 4.3 & 4.2 & \textbf{4.2} \\
        AI Helpfulness (1--5) & 4.5 & 5.0 & \textbf{4.9} \\
        NASA-TLX Mental Demand (1--5) & 3.1 & 3.3 & 3.2 \\
        NASA-TLX Effort (1--5) & 2.9 & 3.2 & 3.0 \\
        NASA-TLX Frustration (1--5) & 2.7 & 2.3 & 2.5 \\
        Confidence pre $\rightarrow$ post & $3.0 \rightarrow 4.1$ & --- & $+1.1$ \\
        \hline
    \end{tabular}
\end{table}

The mean SUS score of 76.4 is above the 68-point benchmark and within
the ``good'' range of the Bangor adjective scale. Engagement of
4.2/5 is slightly above the 4.1/5 reported by Kestin et
al.~\cite{kestin2025aitutor} for a research-designed AI tutor,
although the studies were conducted under different conditions.
Self-reported confidence in identifying and mitigating web
vulnerabilities increased from~3.0 to~4.1 ($+1.1$), with larger gains
among beginners ($+1.6$) than experienced participants ($+0.2$), a
pattern consistent with prior AI-supported learning findings.

The most notable result is the near-identical outcome between the ND
and NT cohorts. ND participants reported marginally higher usability
and engagement, and slightly lower mental demand and effort, than NT
participants, while frustration was the only measure where ND scores
were higher (2.7 vs 2.3). This contrasts with the cognitive-load gap
reported by Le~Cunff et al.~\cite{lecunff2024neurodivergent} for
standard online-learning platforms and is consistent with the design
hypothesis that accessibility-first, scaffolded AI support can reduce
structural learning barriers. On the security pathway, all three
participants who completed the tasks correctly selected mitigations
for CWE-22, CWE-352, and CWE-502, spanning different learner profiles; 
we therefore report this strictly as feasibility evidence. 

\subsection{Qualitative Themes and Limitations}

Thematic analysis of the open-ended responses surfaced five recurring
themes: AI tutor support as the most frequently cited strength;
interface clarity and navigability of the scaffolded ShopSecure
environment; a minority preference for multimodal explanations
(video) alongside text-based tutoring; discoverability gaps for
specific interface elements (subsequently refined mid-pilot); and
limitations in tutor state-awareness across long interactions
(addressed by a four-level prompt redesign during the iterative
cycle). The pilot has obvious limitations: a modest sample
($n=19$) limits statistical generalisation, neurodivergence was
self-reported rather than clinically verified, no control condition
was included, and the qualitative analysis was conducted by a single
coder. Within the Design Science Research framing, these results
provide evidence of design viability rather than population-level
inference.

\section{Conclusion and Future Work}
\label{sec:conclusion}

This paper has presented SYNAPSE, a publicly deployed multi-LLM AI
tutoring platform that combines secure software development
education with a neurodivergent-first interface, coordinated through
the Model Context Protocol. Pilot evidence from nineteen volunteers indicates good usability (SUS~76.4), engagement comparable to recent research-designed AI tutors (4.2/5), and comparable self-reported cognitive-load levels across neurodivergent and neurotypical participants. Future work will expand evaluation, strengthen experimental controls, and extend platform capabilities. 

SYNAPSE is publicly accessible at
\url{https://synapse-course.com}. A short screencast
demonstrating the platform is available at \url{https://youtu.be/9R17KC47qQI}. Source code is available at \url{https://github.com/whitepanther69/synapse-course} and archived at \url{https://doi.org/10.5281/zenodo.20480483}.

\section*{Acknowledgements}

The authors thank the nineteen volunteer participants of the pilot
study. Claude (Anthropic) was used for prose refinement during
manuscript preparation; all technical content and analysis are the
authors' own.
\footnotesize
\bibliographystyle{IEEEtran}
\bibliography{references}

\end{document}